\documentclass[aps,review,superscriptaddress,longbibliography]{revtex4-2}
\usepackage{lineno,hyperref}
\modulolinenumbers[5]

\date{}
\usepackage{booktabs}
\usepackage{siunitx}
\usepackage{amsmath}
\usepackage{subfigure}
\usepackage{float}
\usepackage[capitalize]{cleveref}
\usepackage{xcolor}
\usepackage{amsfonts}
\usepackage{bm}
\usepackage{graphicx}

\usepackage{caption}
\newcommand{\blueblack}{\color{black}}

\newcolumntype{P}[1]{>{\centering\arraybackslash}m{#1}}

\graphicspath{{pic/}}

\begin{document}

\title{Vortex-induced vibration of a flexible pipe under oscillatory sheared flow}

\author{Xuepeng Fu}
\affiliation{State Key Laboratory of Ocean Engineering, Shanghai Jiao Tong University, Shanghai, 200240, China}
\affiliation{Institute of Polar and Ocean Technology, Institute of Marine Equipment, Shanghai Jiao Tong University, Shanghai, 200240, China}

\author{Shixiao Fu}
\email{shixiao.fu@sjtu.edu.cn}
\affiliation{State Key Laboratory of Ocean Engineering, Shanghai Jiao Tong University, Shanghai, 200240, China}
\affiliation{Institute of Polar and Ocean Technology, Institute of Marine Equipment, Shanghai Jiao Tong University, Shanghai, 200240, China}

\author{Mengmeng Zhang}
\affiliation{State Key Laboratory of Ocean Engineering, Shanghai Jiao Tong University, Shanghai, 200240, China}
\affiliation{Institute of Polar and Ocean Technology, Institute of Marine Equipment, Shanghai Jiao Tong University, Shanghai, 200240, China}

\author{Haojie Ren}
\affiliation{State Key Laboratory of Ocean Engineering, Shanghai Jiao Tong University, Shanghai, 200240, China}
\affiliation{Institute of Polar and Ocean Technology, Institute of Marine Equipment, Shanghai Jiao Tong University, Shanghai, 200240, China}

\author{Bing Zhao}
\affiliation{State Key Laboratory of Ocean Engineering, Shanghai Jiao Tong University, Shanghai, 200240, China}
\affiliation{Institute of Polar and Ocean Technology, Institute of Marine Equipment, Shanghai Jiao Tong University, Shanghai, 200240, China}

\author{Yuwang Xu}
\affiliation{State Key Laboratory of Ocean Engineering, Shanghai Jiao Tong University, Shanghai, 200240, China}
\affiliation{Institute of Polar and Ocean Technology, Institute of Marine Equipment, Shanghai Jiao Tong University, Shanghai, 200240, China}

\date{\today}

\begin{abstract}
Vortex-induced vibration (VIV) test of a tensioned flexible pipe in oscillatory sheared flow was performed in an ocean basin. The model was $\SI{28.41}{mm}$ in diameter and $\SI{3.88}{m}$ in length. The test was performed on a rotating test rig to simulate oscillatory sheared flow conditions. One end of the test pipe is fixed, and one end is forced to harmonically oscillate to simulate oscillatory sheared flows with various combinations of amplitudes and periods, Keulegan-Carpenter ($KC$) numbers from $25$ to $160$ and five kinds of reduced velocities $Vr$ from $6$ to $14$. Fiber Bragg Grating (FBG) strain sensors were arranged along the test pipe to measure bending strains, and the modal analysis approach was used to determine the VIV response. The VIV response in the cross flow (CF) direction is investigated. The results show that VIV under oscillatory sheared flow exhibit amplitude modulation and hysteresis phenomena. Compared with oscillatory uniform flow-induced VIV, the Strouhal number is smaller in oscillatory sheared flow-induced VIVs. The VIV developing process in oscillatory sheared flow is analyzed, and critical $KC$ is proposed to describe the occurrence of modulated VIV under oscillatory sheared flow.
\end{abstract}

\maketitle

\section{Introduction}

Vortex-induced vibration (VIV) is a phenomenon that has been extensively studied over the past few decades due to its significant impact on the fatigue damage and drag forces of slender structures such as risers and tension legs in offshore engineering. VIV occurs when fluid flows past these structures, creating a vortex street that induces mechanical oscillations in the in-line (IL) and cross flow (CF) directions. VIV can lead to fatigue damage and ultimately structural failure if not properly accounted for in the design and operation of these structures. A series of studies for VIV have been conducted and can be found in \cite{tognarelli2004viv,williamson2004vortex,sarpkaya2004critical,williamson2008brief,hong2018vortex,duan2021effect,boersma2023experimental}.

Previous studies of VIV have focused primarily on the VIV response of flexible structures under steady flow fields, such as uniform flow \citep{trim2005experimental}, linearly sheared flow \citep{song2016distribution}, stepped flow \citep{chaplin2005laboratory}, and inclined uniform flow \citep{han2017dynamic}. One of the detailed studies regarding the VIV flexible pipes is the Norwegian Deepwater Program (NDP) joint industrial project (JIP). NDP \citep{vikestad2000norwegian} has performed a series of VIV experiments on flexible risers in uniform flow and linearly sheared flow.  Many empirical prediction models have been proposed and revised based on experimental results, such as SHEAR7 \citep{vandiver2005user}, VIVANA \citep{larsen2001vivana} and VIVA \citep{triantafyllou1999pragmatic}.

However, flexible slender structures in actual ocean engineering always suffer from spatiotemporally varying flow fields. \citet{wang2015out} conducted a VIV experiment of a steel catenary riser (SCR) under a vessel motion-induced complex flow field. \citet{wang2015fatigue} simplified such a complex spatiotemporally varying flow field to oscillatory uniform flow and investigated the VIV characteristics of flexible cylinders under oscillatory uniform flow \cite{fu2014features}. However, to the best of our knowledge, there have been no VIV experiments on slender pipes in oscillatory sheared flow.

\begin{figure}[h]
	\centering
	\includegraphics[width=.4\textwidth]{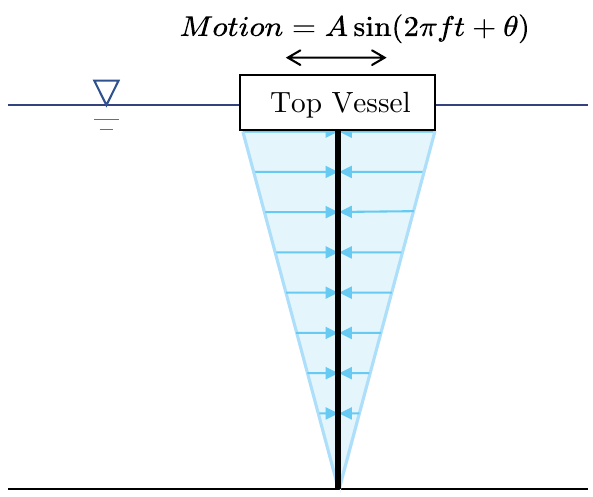}
	\caption{Sketch of the top vessel motion-induced oscillatory sheared flow. The top vessel motion is assumed to be a harmonic motion.}
	\label{picflowdia}
\end{figure}

Oscillatory sheared flow is an important flow field to consider when designing and operating offshore structures. Top tension risers (TTRs), for example, are often used to connect a floating production platform to the seabed. One end of the TTR is fixed to the platform, while the other end is forced to move with the top vessel motion, which generates an oscillatory sheared flow, as shown in \cref{picflowdia}. This flow field can significantly impact the VIV response of the TTR, but little is currently known regarding the VIV characteristics of slender pipes in oscillatory sheared flow.

In the present study, we aim to fill this gap in knowledge by performing the VIV experiment of a flexible pipe in oscillatory sheared flow. The model used in this study has a diameter of $\SI{28.41}{mm}$ and a length of 137 diameters. This study will provide important insights and benchmark data into the VIV characteristics of slender pipes in oscillatory sheared flow and help to improve the design and operation of offshore structures.

\section{Experimental setup}
\subsection{Test apparatus}

\begin{figure}[h]
	\centering
	\includegraphics[width=.7\textwidth]{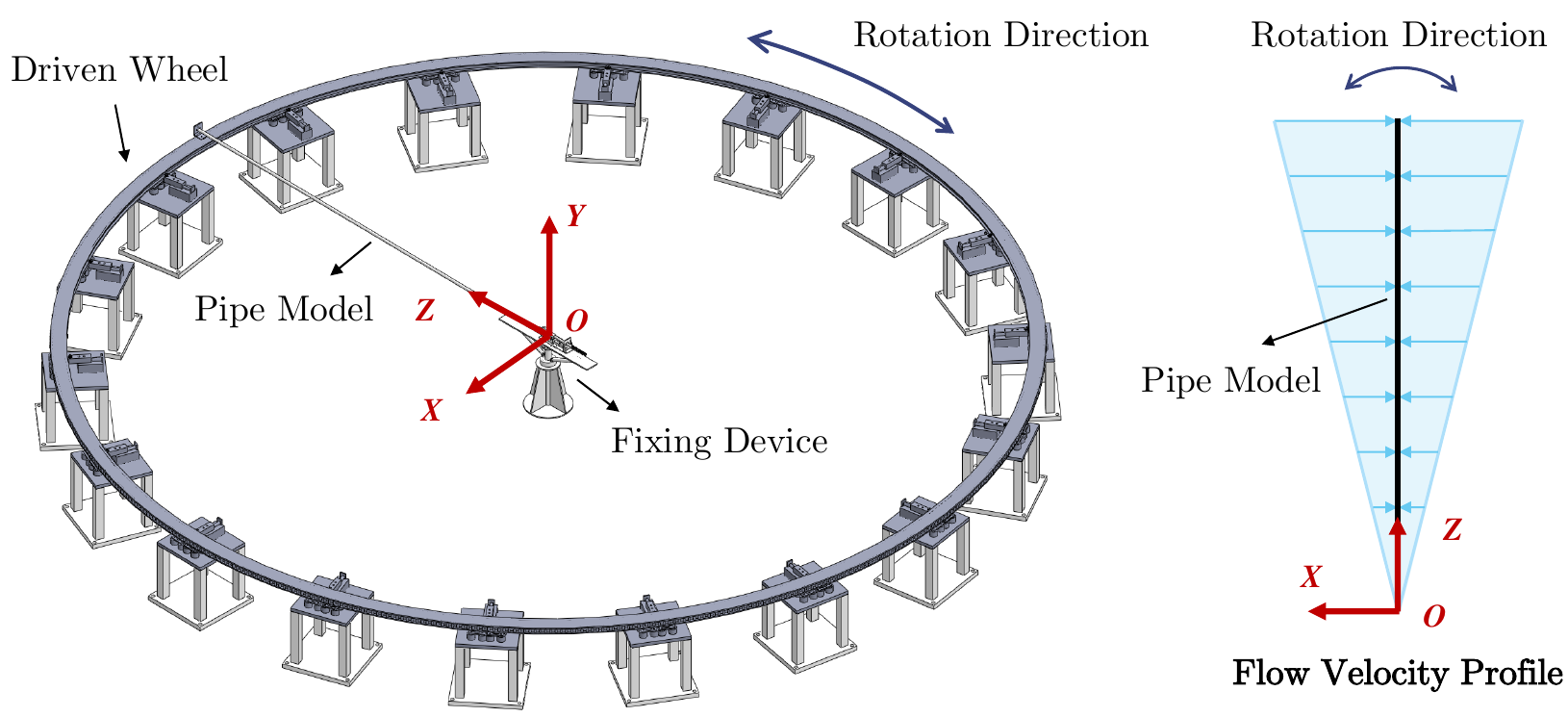}
	\caption{Sketch of the experimental apparatus. Left: experimental apparatus with a rotating wheel driven by a servo motor; Right: effective flow velocity profile.}
	\label{pic2}
\end{figure}

The model test was performed in the ocean basin of the State Key Laboratory of Ocean Engineering at Shanghai Jiao Tong University. The experimental apparatus, which includes the rotating rig and the tested pipe model, is installed on the false bottom of the basin, as shown in \cref{pic2}. The experimental apparatus produced the flow field by rotating through a timing belt motivated by the servo motor. This novel VIV experimental apparatus has been validated through credibility analysis, including noise signal analysis, repetitive experiments and water depth tests, as shown in previously published research \citep{fu2022experimental,fu2022study}. In contrast to the experiment of VIV under bidirectionally sheared flow, we set a fixing device in the center of the test apparatus. The test pipe model was installed on two sides of the edge of the test apparatus (driven wheel) and fixing device, which was equipped with clamps, U-joints and a force sensor. A pretension force of $\SI{550}{N}$ was applied to the pipe model through the tensioner.

The coordinate system was defined as \emph{O-XYZ}, as shown in \cref{pic2}. The origin ($O$) was set at one ending point of the pipe model connected with a fixing device. The \emph{X-}, \emph{Y-} and \emph{Z-} axes were along the in-line (IL), cross-flow (CF) and axis of the test pipe model. The rotation direction during the experiment was always along the \emph{X-}axis, and the effective current profile is shown with colors and arrows.

The primary physical properties of the test pipe model are listed in \cref{tab1}. The first natural frequencies were obtained by a free decay experiment, and the calculated 1st natural frequency in water was calculated by \cref{eq1}. The theoretical value of the 1st natural frequency differed from the experimentally measured value by 5\%.

\begin{table}[h]
	\centering
    \caption{Physical properties of the test pipe model.}
	\label{tab1}
	\begin{tabular}{@{}ll@{}}
		\toprule
		Parameter                                                                 & Value of test model \\ \midrule
		Pipe model length $L$ [m]                                              & 3.88  \\
		Outer diameter $D$ [mm]                                                & 28.41 \\
		Mass in air $\bar{m}$ [kg/m]                          & 1.24  \\
		Pretension  $T_{0}$ [N]                                                      & 490   \\
		Bending stiffness $EI$ [$\SI{}{Nm^2}$] & 58.6  \\
		Tensile stiffness $EA$ [N]                                             & 9.4E5 \\
		Damping ratio $\zeta$ [\%]                             & 2.58  \\
          mass ratio $m^*$ [-]   &   1.96\\
		Tested 1st natural frequency in water $f_{n1}$ [Hz]                      & 2.29  \\
		Calculated 1st natural frequency in water $f_{cn1}$ [Hz]                   & 2.20  \\ \bottomrule
	\end{tabular}
\end{table}

\begin{equation}\label{eq1}
f_{c n}=\frac{n^{2} \pi}{2} \sqrt{\left(1+\frac{L^{2} T_{0}}{n^{2} \pi^{2} E I}\right) \frac{E I}{\bar{m}^{\prime} L^{4}}},\quad \bar{m}^{\prime}=\bar{m}+\frac{1}{4} C_{m} \rho \pi D^{2}\left(C_{m}=1.00\right) \quad(n=1,2,3 \ldots),
\end{equation}
where $T_0$ is the pretension, and $\bar{m}$ and $\bar{m}^{\prime} $ are the mass per unit length in air and still water, respectively. $D,L$ and $EI$ are the diameter, length and bending stiffness of the test pipe model, respectively. $\rho$ is the density of water, $\rho=\SI{1000}{kg/m^3}$, and the added mass coefficient $C_m$ was set to 1.00 \citep{khalak1997investigation}. The mass ratio is designed based on a real Steel Catenary Riser, and the tension is given to ensure the pipe model is tension dominated.

\begin{figure}[ht!]
	\centering
	\includegraphics[width=.8\textwidth]{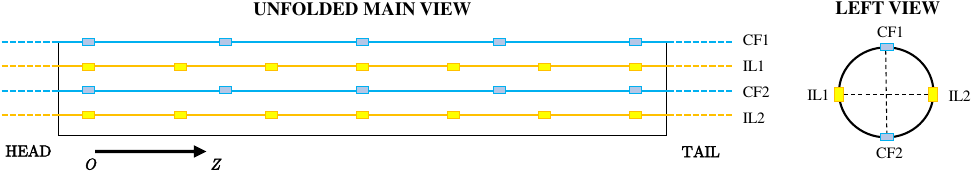}
	\caption{Locations of the FBG strain sensors along the test pipe model.}
	\label{picFBG}
\end{figure}

A total of 24 Fiber Bragg Grating (FBG) strain sensors in four groups are installed along the test pipe to measure dynamic strains in both the IL and CF directions, as shown in \cref{picFBG}. There are five strain gauge points distributed in the CF direction and seven strain gauge points distributed in the IL direction. The strains along the pipe model were synchronously recorded at a sampling frequency of $\SI{250}{Hz}$.

\subsection{Test condition}
The pipe model is forced to perform oscillatory motion during the experiment to simulate the oscillatory sheared flow. In contrast to the oscillatory uniform flow test, the motion along the pipe model is varied in the present study. We define the motion of the endpoint ($Z=L$) with the maximum flow velocity as:
\begin{align}
    X_m(t)&=A_m\sin(2\pi f_o t),\\
    U_m(t)&=\dot{X}_m(t)=2\pi f_0 A_m\sin(2\pi f_o t),
\end{align}
where $X_m(t)$ is the motion of the endpoint, $U_m(t)$ is the velocity of the forced motion, $A_m$ is the oscillation amplitude and $f_o$ is the oscillation frequency. The oscillation period $T$ is defined as ${1}/{f_o}$. The Keulegan–Carpenter number ($KC$), reduced velocity ($Vr$) and maximum Reynolds number ($Re_{max}$) are involved in the present study and defined as:
\begin{align}
    &KC=\frac{2\pi A_m}{D},\\
    &Vr = \frac{2\pi f_o A_m}{f_{cn1}D},\\
    &Re_{max} = \frac{U_{max}}{\nu D},
\end{align}
where $f_{cn1}$ is the calculated 1st natural frequency, as shown in \cref{tab1}, $\nu = 1.00\times 10^{-6}$ is the kinematic viscosity of fluid. Five kinds of reduced velocities $Vr = 6,8,10,12$ and $14$ with $KC$ varying from $20$ to $160$ are carried out. The corresponding maximum Reynolds number ($Re_{max}$) are $10654$, $14205$, $17757$, $21308$ and $24860$, respectively.

\section{Data analysis}
\subsection{Preprocessing}
 The measured strain at the sensor location in the CF direction $\varepsilon_{CF}\left(z,t\right)$ contains two parts: the variable axial strain $\varepsilon_T\left(t\right)$ caused by varying tension, and the bending strain $\varepsilon_{VIV}\left(z,t\right)$ caused by VIV. The strain at $z$ locations of CF1 and CF2 can therefore be written as:
\begin{align}
&\varepsilon_{CF1}(z, t)=\varepsilon_{T}(t)+\varepsilon_{C F-VIV}(z, t), \label{eq2}\\
&\varepsilon_{CF2}(z, t)=\varepsilon_{T}(t)-\varepsilon_{C F-VIV}(z, t). \label{eq3}
\end{align}

\noindent Combining \cref{eq2} and \eqref{eq3}, the CF VIV strain at the $z$ location can be calculated by: 
\begin{equation}
\varepsilon_{C F-VIV}(z, t)=\left[\varepsilon_{C F 1}(z, t)-\varepsilon_{C F 2}(z, t)\right] / 2.
\end{equation}

\noindent Since the VIV in the IL direction is a periodic process with a mean value of zero, the time-averaged value of the IL-VIV strain equals zero:
\begin{equation}
\overline{\varepsilon_{I L-V I V}(z, t)}=0.
\end{equation}

\noindent The IL VIV strain can be derived as:
\begin{align}
	\varepsilon_{initial }(z)&=\left[\overline{\varepsilon_{I L 1}(z, t)}-\overline{\varepsilon_{I L 2}(z, t)}\right] / 2, \\
	\varepsilon_{IL-VIV}(z, t)&=\left[\varepsilon_{I L 1}(z, t)-\varepsilon_{I L 2}(z, t)-\overline{\varepsilon_{I L 1}(z, t)}+\overline{\varepsilon_{I L 2}(z, t)}\right] / 2.
\end{align}

\noindent Bandpass filtering was applied to the strain signal to eliminate unavoidable noise \citep{lie2006modal}. The high-pass and low-pass cutoffs were $\SI{0.8}{Hz}$ and $\SI{20}{Hz}$, respectively.

\subsection{Modal analysis method}
The VIV displacement in the IL and CF directions can be obtained by the modal analysis method through the measured IL and CF VIV strain.

\begin{figure}[!h]
	\centering
	\includegraphics[width=.6\textwidth]{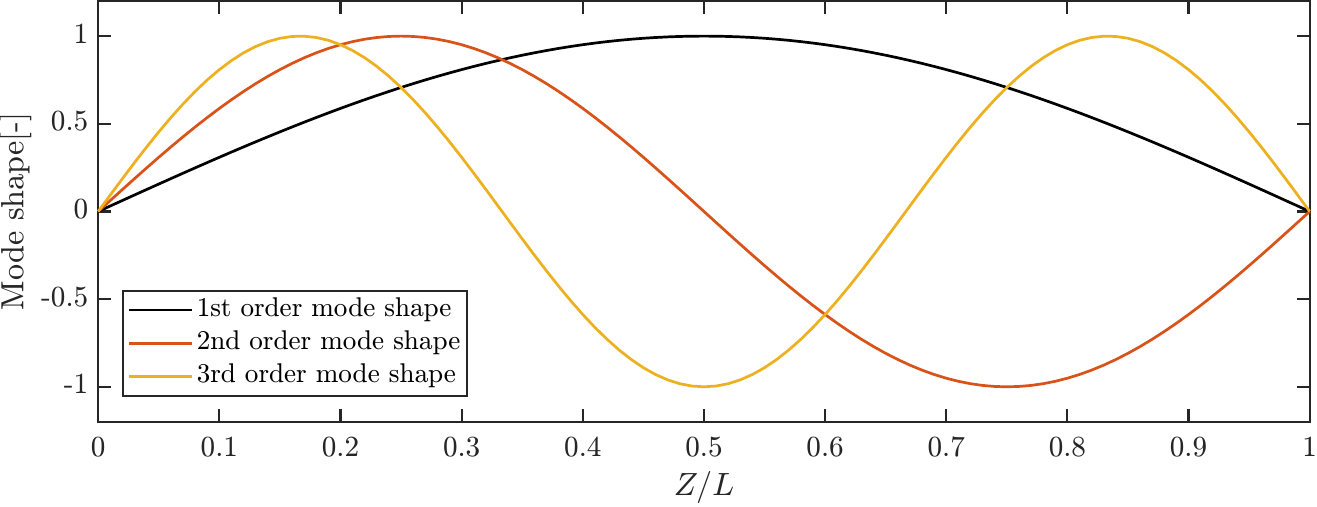}
	\caption{First three vibration mode $\varphi(z)$ of the pipe model in the present study.}
	\label{picmode}
\end{figure}

The modal analysis method assumed that the VIV displacement can be expressed as a sum of the modal shapes multiplied by the modal weights at each time step. The VIV displacement in the CF direction can be expressed as:
\begin{equation}\label{eq12}
y(z, t)=\sum_{i=1}^{n} p_{i}(t) \varphi_{i}(z), \quad z \in[0, L],
\end{equation}
where $y\left(z,t\right)$ is the VIV displacement in the CF direction at the $z$ location, $p_i\left(t\right)$ is the $i$th modal weight of the displacement at time $t$, and $\varphi_i(z)$ is the $i$th modal shape of the displacement at the $z$ location. \cref{picmode} displays the first three modes of the pipe model in the present manuscript.

Based on the small deformation assumption, the curvature $\kappa\left(z,t\right)$ can be derived from the displacement modal shape as:
\begin{equation}
\kappa(z, t)=\frac{\partial^{2} y(z, t)}{\partial z^{2}}=\sum_{i=1}^{n} p_{i}(t) \varphi_{i}^{\prime \prime}(z), \quad z \in[0, L],
\end{equation}
where $\varphi^{\prime\prime}\left(z\right)$ is the $i$th modal weight of the curvature. Based on the relationship between curvature and strain, we have:
\begin{equation}\label{eq14}
\varepsilon(z, t)=\kappa(z, t) R=R \sum_{i=1}^{n} p_{i}(t) \varphi_{i}^{\prime \prime}(z), \quad z \in[0, L],
\end{equation}
where $R=D/2$ is the radius of the pipe model.

For the test pipe model, the modal shapes of the displacement can be treated as the modal shape of the beam pinned at both ends, which can be expressed as:
\begin{equation}
\varphi_{i}(z)=\sin \frac{i \pi z}{L}, \quad i=1,2 \cdots.
\end{equation}

\begin{figure}[!th]
	\centering
	\includegraphics[width=.6\textwidth]{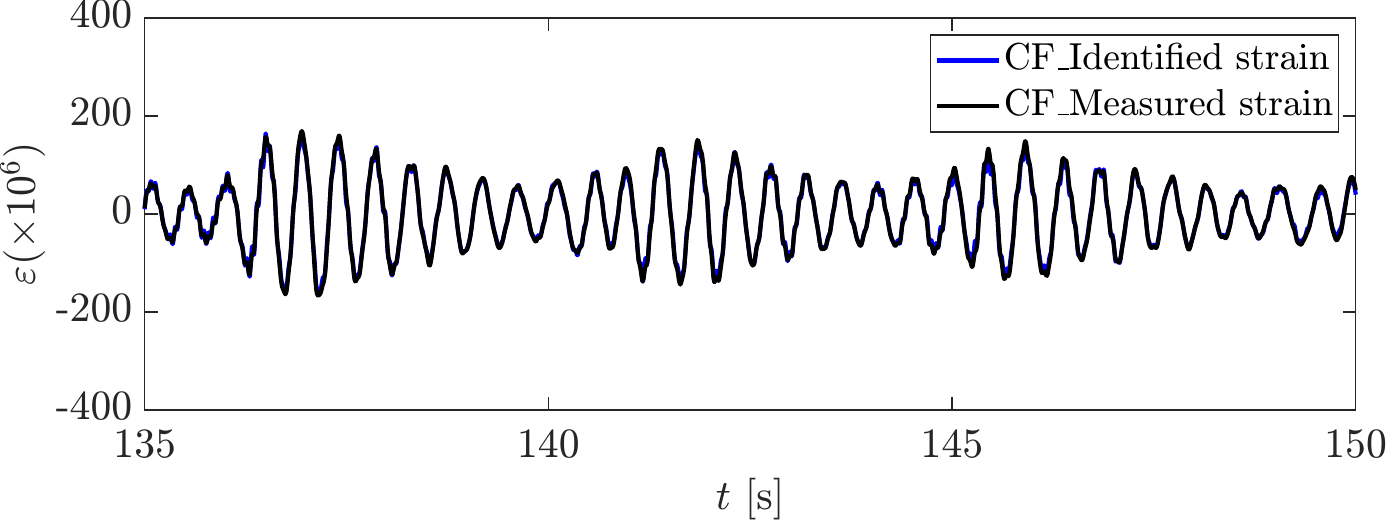}
	\caption{Comparison between the identified and measured strain in the case $Vr=8$, $KC=160$ at FBG-CF3.}
	\label{picmodevalida}
\end{figure}

\noindent The modal shapes of the curvature $\varphi^{\prime\prime}\left(z\right)$ are also sinusoidal, and the strain can be calculated by:
\begin{equation}\label{eq16}
\varepsilon(z, t)=-R \sum_{i=1}^{n}\left(\frac{i \pi}{l}\right)^{2} p_{i}(t) \varphi_{i}(z), \quad z \in[0, L],
\end{equation}
where $\varepsilon\left(z,t\right)$ is the VIV bending strain. Then, the modal weight of the displacement $p_i\left(t\right)$ can be obtained through \cref{eq16}, and the VIV displacement in the CF direction can be derived directly using \cref{eq12}. The derivation of VIV displacement in the IL direction $x\left(z,t\right)$ follows similarly.  \cref{picmodevalida} shows the comparison result of the identified and measured strain in the case $Vr=8$, $KC=160$ at FBG-CF3. The identified strain is obtained by the second order difference, the results show the identified CF strains are in good agreement with the measured strain in the time domain.

\subsection{Time-frequency analysis}
 
To investigate the time-frequency feature of the VIV response under bidirectionally sheared flow, wavelet transformation is used. The continuous wavelet transform equation is expressed as:
\begin{equation}
WT_{f}(a, \tau)=\left\langle f(t), \psi_{a, \tau}(t)\right\rangle=a^{-1 / 2} \int_{-\infty}^{+\infty} f(t) \psi^{*}\left(\frac{t-\tau}{a}\right) d t,
\end{equation}
where $WT_{f}$ is the wavelet transformation coefficient of the time domain signal $f(t)$, which represents the variation in frequency at that time scale. Parameter $a$ is the scale factor, $\tau$ is the shift factor, $\psi(t)$ is the mother wavelet, and the Morlet wavelet is chosen as the mother wavelet. Besides wavelet transform, Fast Fourier Transform (FFT) is applied for determing the frequency components of the VIV response of the pipe \citep{thorsen2016time,barbi1986vortex,opinel2020application}.

\section{Results and discussion}
\subsection{VIV responses in the oscillatory sheared flow}
\begin{figure}[ht!]
	\centering
	\includegraphics[width=.7\textwidth]{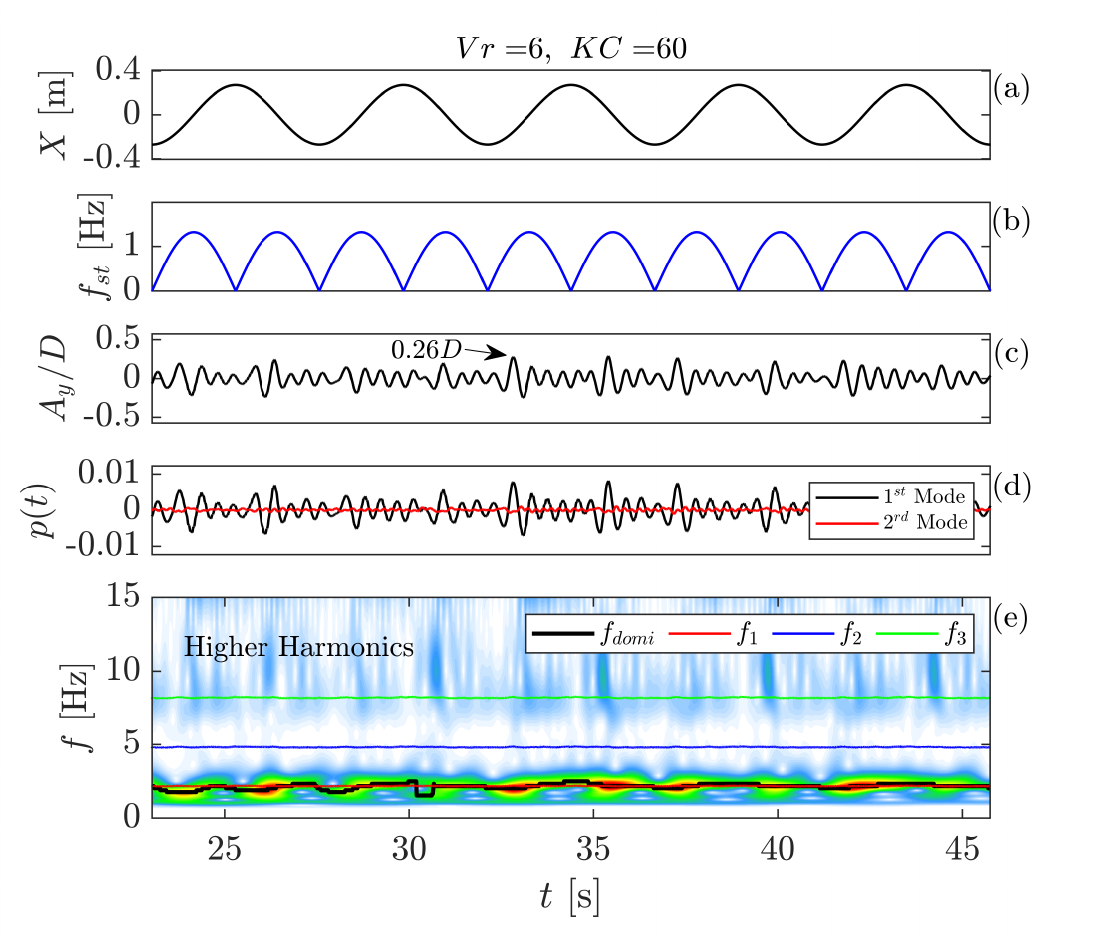}
	\caption{VIV response in case $Vr=6, KC=60$ at FBG-CF3. From top to bottom: (a) displacement of the edge; (b) theoretical time varying shedding frequency; (c) normalized time-varying VIV displacement; (d) modal weight of the first and second mode; black line: the first modal weight, red line: the second modal weight and (e) time-varying response frequency by wavelet analysis; black line: dominant frequency by wavelet analysis, red line: instantaneous natural frequency of the first mode, blue line: second mode, green line: second mode.}
	\label{vr6kc60time}
\end{figure}

\cref{vr6kc60time} shows the VIV response at the case with $Vr=6, KC=60$ at FBG3-CF with five subfigures: (a) displacement of the edge measured by the encoder during the experiment; (b) theoretical time varying shedding frequency calculated by:
\begin{equation}
    f_{st}=\dfrac{St U_g}{D},
\end{equation}
where $St=0.20$, $U_g$ is the calculated relative flow velocity of the gauge point and $D$ is the diameter of the tested pipe; (c) normalized time-varying VIV displacement in the CF direction by modal analysis; (d) modal weight of the first and second modes calculated by modal analysis; (e) time-varying VIV frequency by wavelet analysis and instantaneous natural frequency in the first three modes based on measured tension calculated by \cref{eq1}.

\begin{figure}[!ht]
	\centering
	\includegraphics[width=.55\textwidth]{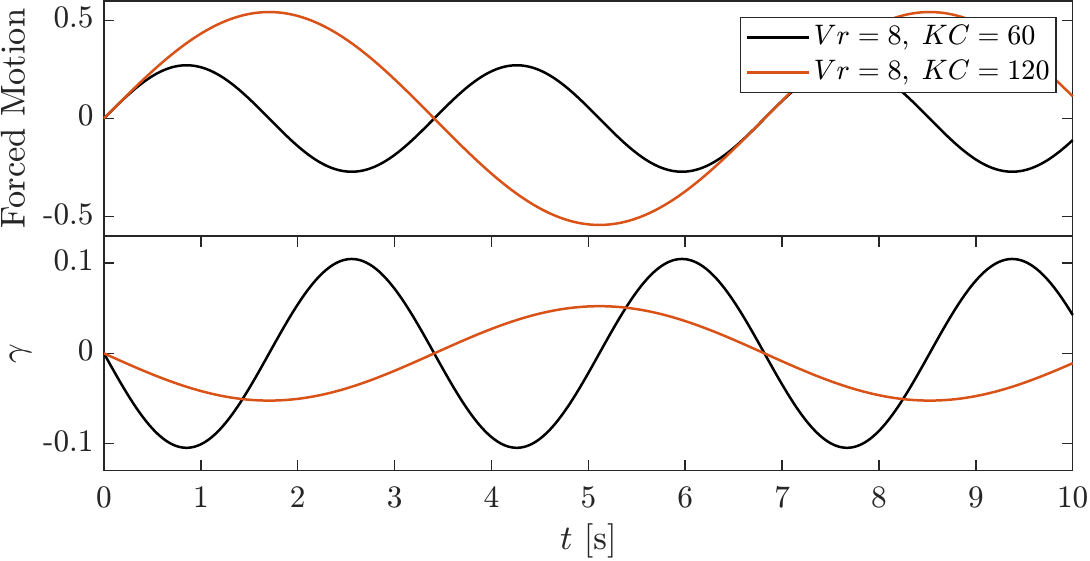}
	\caption{Forced motion and corresponding $\gamma$ in different $KC$ case (same $Vr$).}
	\label{picgamma}
\end{figure}

\begin{figure}[ht!]
	\centering
	\includegraphics[width=.7\textwidth]{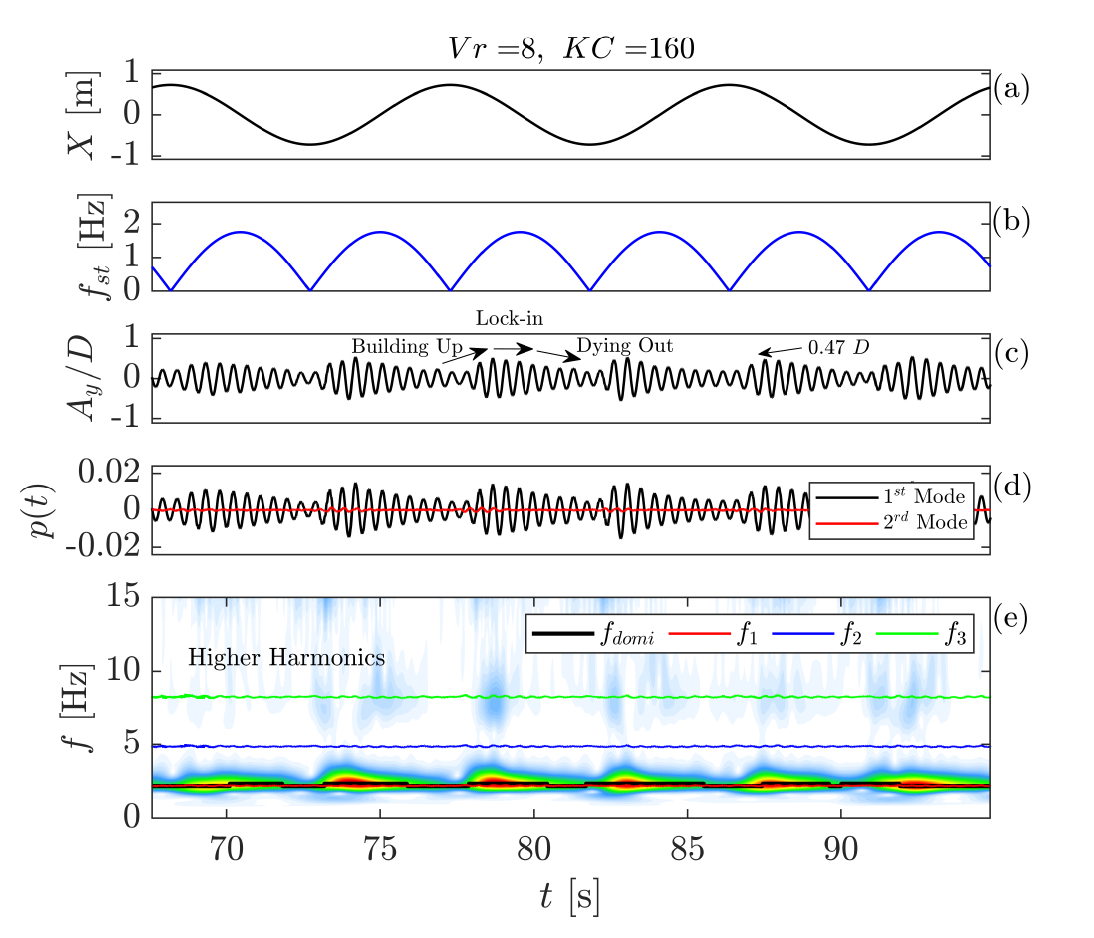}
	\caption{VIV response in case $Vr=8, KC=160$ at FBG-CF3. From top to bottom: (a) displacement of the edge; (b) theoretical time varying shedding frequency; (c) normalized time-varying VIV displacement; (d) modal weight of the first and second mode and (e) time-varying response frequency by wavelet analysis.}
	\label{vr8kc160time}
\end{figure}

\cref{vr6kc60time} shows that the VIV response in small $Vr$ and $KC$ is relatively stable with a maximum response amplitude of $0.26D$. The VIV response is dominated by the first mode with higher harmonics. \cref{vr8kc160time} displays the VIV response under the case with $Vr=8, KC=160$, and it can be seen that an obvious modulated VIV response occurs with a maximum response amplitude of $0.47 D$. The modulated VIV response is divided into three regions: building up region, lock-in region and dying out region. There also exists an obvious higher harmonic in this case. Compared with the low $Vr$ and $KC$ cases, the modulated vibration is more significant, and the vibration is dominated by the first order in the present experiment. $\gamma$ parameter defined as 
\begin{equation}
    \gamma = \frac{\partial U}{\partial t}\frac{D}{U_{max}^2}
\end{equation}
is proposed to evaluate the rate of flow velocity change, where $U$ is the time-varying velocity, $D$ is the diameter and $U_{max}$ is the maximum velocity. \cref{picgamma} displayed the forced motion and corresponding $\gamma$ in case $Vr=60$ and $KC=120$ with same $Vr=8$. The black line (small $KC$) represents a low motion amplitude but a high $\gamma$, which means the flow velocity change more rapidly. Therefore, the velocity change rapidly in small $Vr$ and $KC$ cases where the flexible pipe cannot react quickly enough to build up VIV response, which results in a stable response. More research on the mechanism of $\gamma$ will be conducted in the future.

The time-varying response results of the oscillatory sheared flow induced VIVs are nearly the same as those of linearly oscillatory flow \citep{wang2015fatigue}.

\subsection{Hysteresis}
\begin{figure}[ht!]
	\centering
	\includegraphics[width=.65\textwidth]{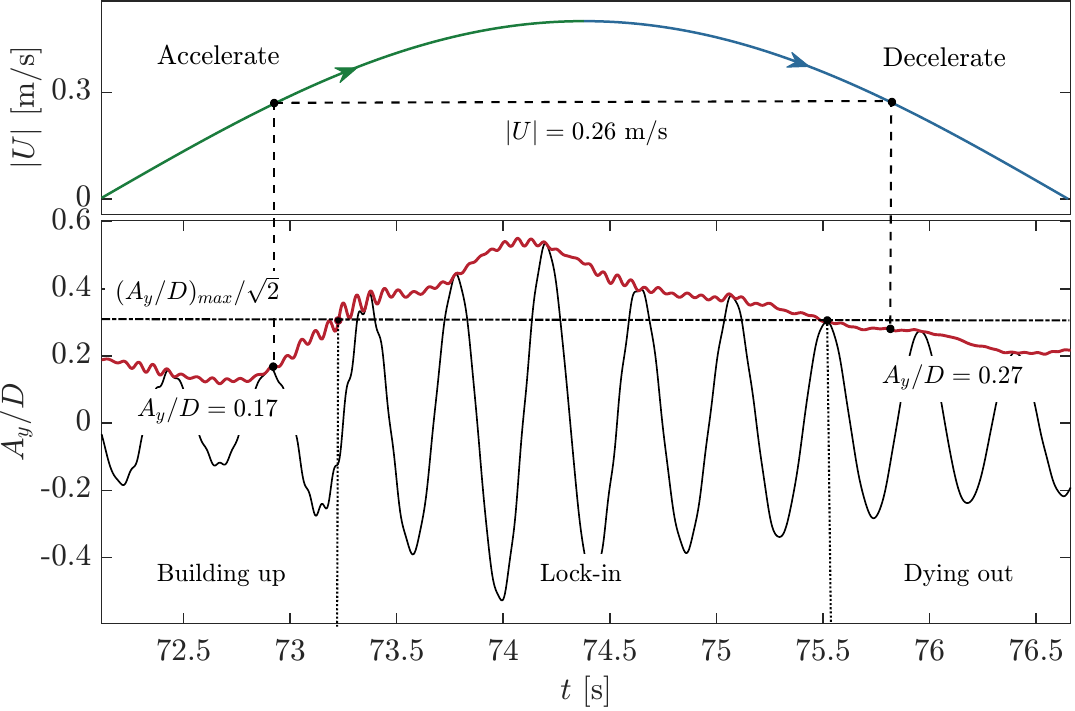}
	\caption{VIV response in case $Vr=8, KC=160$ at FBG-CF3 in the half-period. From top to bottom: absolute value of relative flow velocity at the gauge point of the half period; normalized time-varying VIV displacement. Green line: acceleration region, blue line: deceleration region, black line: time-varying VIV response, red line: instantaneous displacement envelope by Hilbert transform, dotted line: discriminate line for three regions with the value of $(A_y/D)_{max}/\sqrt{2}$.}
	\label{timevr8kc160}
\end{figure}

Hysteresis is a phenomenon discovered in bluff cylinder VIV systems in which the VIV response is inconsistent in the acceleration and deceleration regions \citep{jauvtis2004effect}. \cref{timevr8kc160} displays the time-varying response in the case with $Vr=8$ and $KC=160$ at the FBG-CF3 gauge point. The green and blue lines represent the acceleration and deceleration regions in the half-motion period, respectively. The black and red lines represent the normalized time-varying VIV displacement and instantaneous displacement envelope, respectively. The distribution of the VIV displacements in the acceleration and deceleration regions is not entirely symmetrical. When $|U|= \SI{0.26}{m/s}$, $A_y/D$ is $0.17$ in the acceleration region and $0.27$ in the deceleration region. The absolute value of the gradient of the response amplitude increase in the acceleration region is larger than that in the deceleration region. Moreover, the dotted line with a value of $(A_y/D)_{max}/\sqrt{2}$ separates the response into three regions named the building up, lock-in and dying out regions to describe the developing process.

\begin{figure}[ht!]
	\centering
	\includegraphics[width=.7\textwidth]{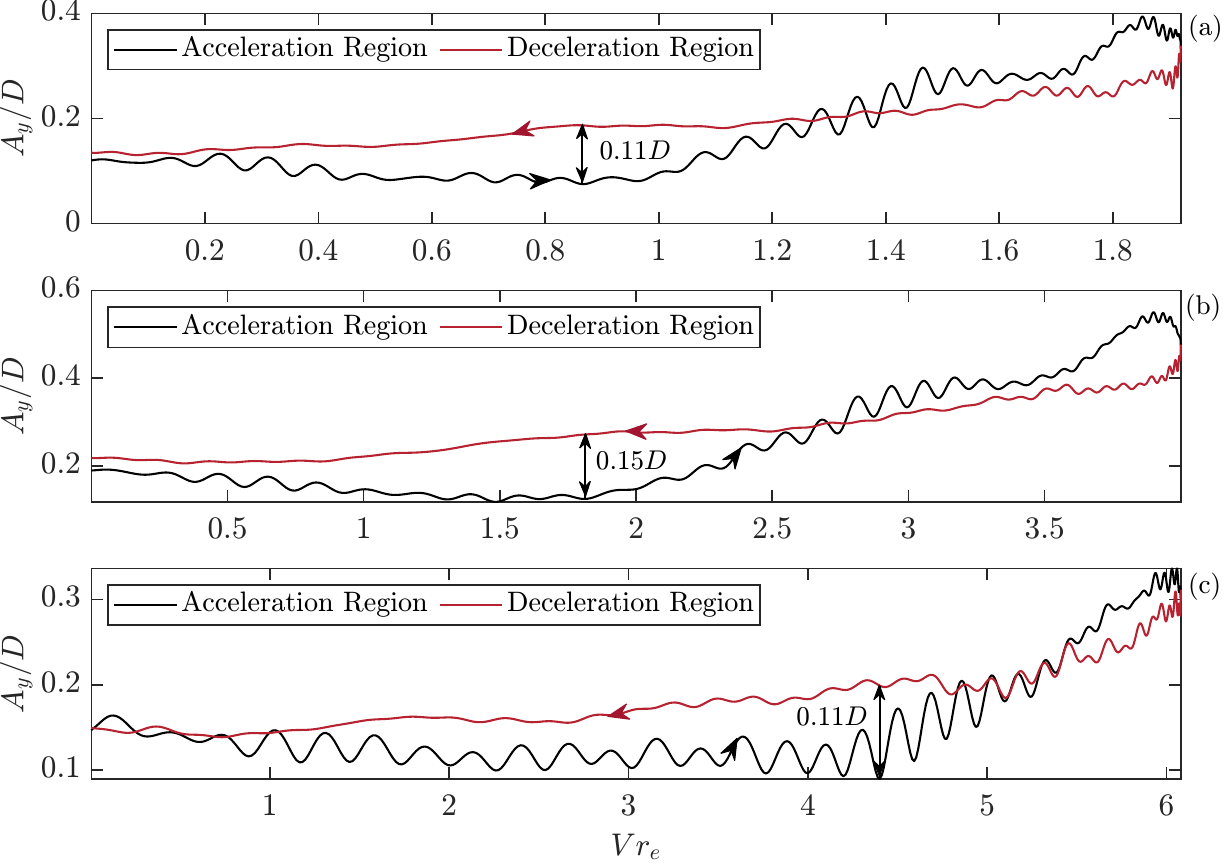}
	\caption{Hysteresis process in case $Vr=8, KC=160$ at different gauge points. Red line: VIV response in the deceleration region; Black line: VIV response in the deceleration region. (a) hysteresis process in FBG-CF1; (a) hysteresis process in FBG-CF3; (a) hysteresis process in FBG-CF5.}
	\label{hystevr8kc160}
\end{figure}

\cref{hystevr8kc160} further demonstrates the response amplitude in the acceleration and deceleration regions with the same effective reduced velocity $Vr_e$ defined as ${|U|}/{f_{cn1}D}$. $|U|$ is the instantaneous effective flow velocity, as \cref{timevr8kc160} shows. The black and red lines represent the response amplitude in the acceleration and deceleration regions, respectively. The arrows depict the $Vr_e$ variation in half a cycle. There exists a difference in the response amplitude in the acceleration and deceleration regions. The response in the deceleration is larger than that in the acceleration region at low $Vr_e$. Combining our previous experimental results on VIVs under uniform oscillatory flow, the hysteresis phenomenon in oscillatory flow-induced VIVs can be defined as \textit{smaller gradients in the VIV response in the deceleration region compared to the acceleration region.} Hydrodynamic analysis shows that the flexible pipe extracts more energy from the fluid to support the vibration which resulted in a higher gradient as Appendix A shows.

\subsection{Strouhal number}
{\blueblack
\cref{diffrentVrtime} displays the response frequencies at different locations of different cases. The displacement of the edge and time-varying response frequency at $Z = L/6$, $L/2$ and $5L/6$ are shown from top to bottom. The black line on the right is the corresponding FFT result based on the time-domain signal. The result shows that the dominant frequencies are consistent in time domain and there exists more harmonic signal in the small $Vr$ and $KC$ case. The results are similar with \cref{vr6kc60time} and \cref{vr8kc160time}. The FFT result of small $Vr$ and $KC$ case presents more frequency components than large $Vr$ and $KC$ case, and the frequency components are basically the same along the cylinder.

The dominant response frequency is defined in the present study as:
\begin{equation}
    f_{domi}=\mathop{\arg\max}_{f} \mathcal{F}(f)=\mathop{\arg\max}_{f} \sum_{i=1}^n \hat{\mathcal{F}}_i(f),
\end{equation}
where $f_{domi}$ is the dominant response frequency, $\mathcal{F}(f)$ is the general frequency spectrum defined as the sum of the power spectral density of the measured signal at the $i$th strain gauge point $\hat{\mathcal{F}}_i(f)$, and $n$ is the number of strain gauges, which is five for the CF direction in the present study. It should be noted that FFT method is not the optimal approach to determine the dominant frequency for time-varying signal. The FFT and wavelet transform results are similar and basically the same along the pipe, therefore FFT is employed in the present study to obtain the dominant response frequency.

\begin{figure}[ht!]
\centering
\includegraphics[width=.95\textwidth]{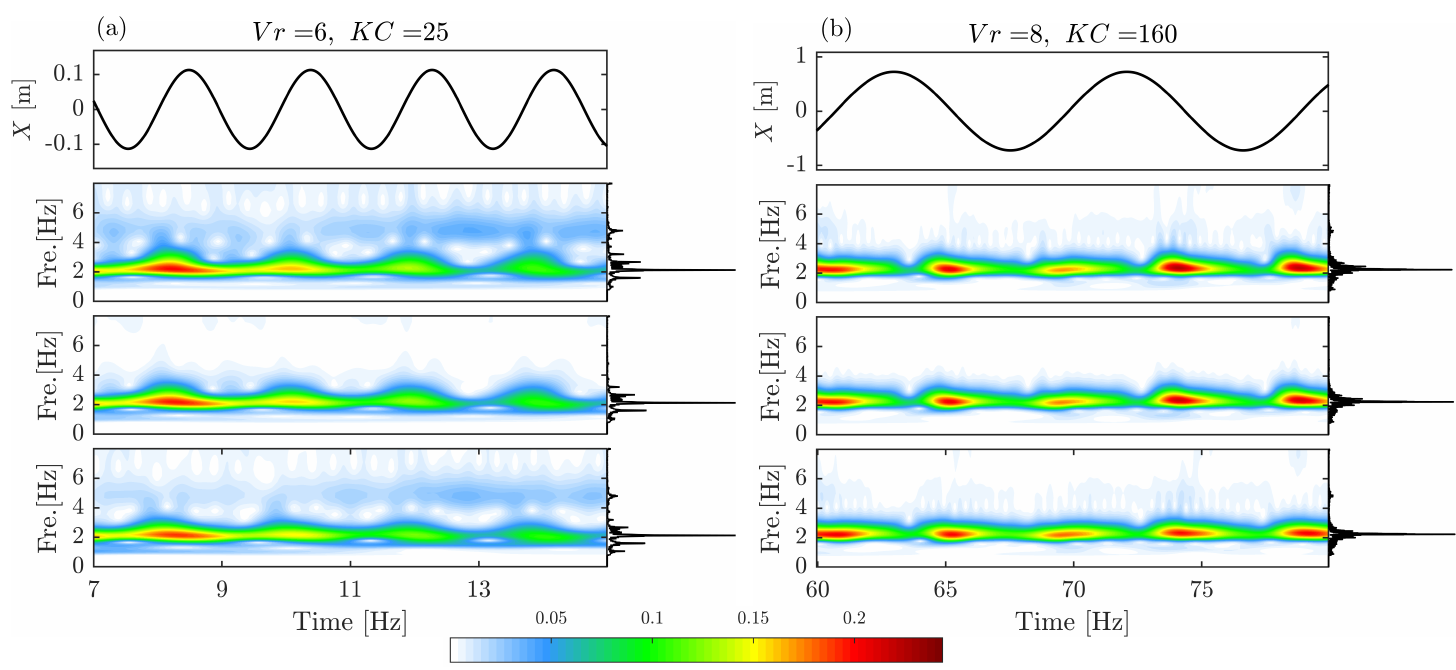}
\caption{Response frequencies at different locations of different cases. From top to bottom: displacement of the edge; time-varying response frequency at $Z = L/6$, $L/2$ and $5L/6$, respectively. Black line on the right represents the corresponding FFT results. Subfigure (a): $Vr=6$, $KC=25$ case, subfigure (b): $Vr=8$, $KC=160$ case.}
\label{diffrentVrtime}
\end{figure}

\cref{FFTre} represents the general frequency spectrum $\mathcal{F}(f)$ in the cases with $Vr=6, KC=25$ and $Vr=8, KC=160$. In the low $KC$ cases, the dominant frequency always lock in the multi-frequency response close to the natural frequency. The dominant frequency in the case of $KC=25$ is $\SI{2.13}{Hz}$ ($4f_o$). The higher harmonic response, which includes the 2nd, 3rd, 5th, 6th, and 9th harmonics, is also involved in the VIV response. For cases with large $KC$, broadband spectra with a dominant frequency close to the natural frequency exist.

\begin{figure}[ht!]
	\centering
	\includegraphics[width=.8\textwidth]{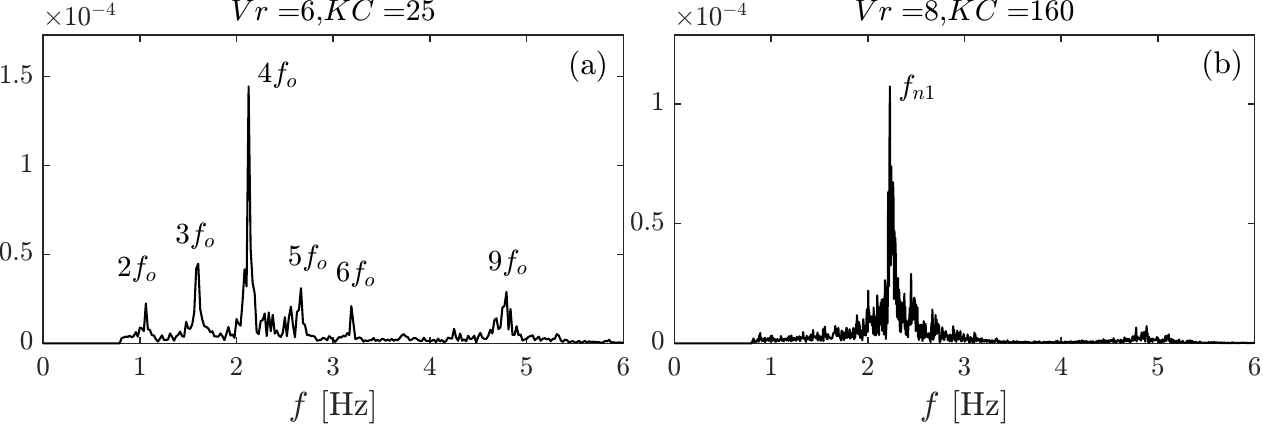}
	\caption{General frequency spectrum of case (a) $Vr=6, KC=25$, (b) $Vr=8, KC=160$. }
	\label{FFTre}
\end{figure}
}

{\blueblack 
The number of vortex shedding (frequency ratio) $N$ is defined as:
\begin{equation}
    N=\dfrac{f_{domi}}{f_o},
\end{equation}
where $f_o$ is the frequency of oscillation. Then we have the Strouhal number $St$ expressed as:
\begin{equation}
    St = \frac{N}{KC} = \frac{f_{domi} D}{U_m},
\end{equation}
The Strouhal number $St$ reflects the VIV excitation effect of the flow field. A larger $St$ indicates that the flow field can potentially excite a higher VIV response frequency. The Strouhal number $St$ of the VIV response under uniform flow is chosen as $0.18$. It should be noted that the $St$ is used for the vortex shedding frequency prediction first, and it is used for the description of slender structure vibration frequency in steady flow later \citep{lie2006modal} (also nondimensional frequency \citep{thorsen2016time}). In the present study, the background flow velocity varies with time, and the $St$ is not stable, leading to a unstable response frequency \citep{dahl2010dual,dahl2006two}. The $St$ is estimated by the dominant frequency obtained by FFT, and the applicability of proposed $St$ for VIV prediction in unsteady flow will be investigated further.

\cref{NversusKC} displays the number of vortex shedding $N$ versus $KC$ in different cases. The black circles represent the result obtained from the experiment, the black line represents the fitted result based on the scatter experimental results, and the blue line represents the result with $St=0.13$ which is proposed for the linearly sheared flow induced VIV frequency prediction \citep{gao2015viv}. $St$ is approximately $0.17$ in the case with $Vr=6$ and $0.13$ in the case with $Vr=8$ and $0.10$ and in the case with $Vr=10$. For the case of $Vr = 8$, the $St$ is basically the same with the linearly sheared flow case, the $St$ decreases with the $Vr$ increased.

\begin{figure}[ht!]
	\centering
 	\includegraphics[width=.6\textwidth]{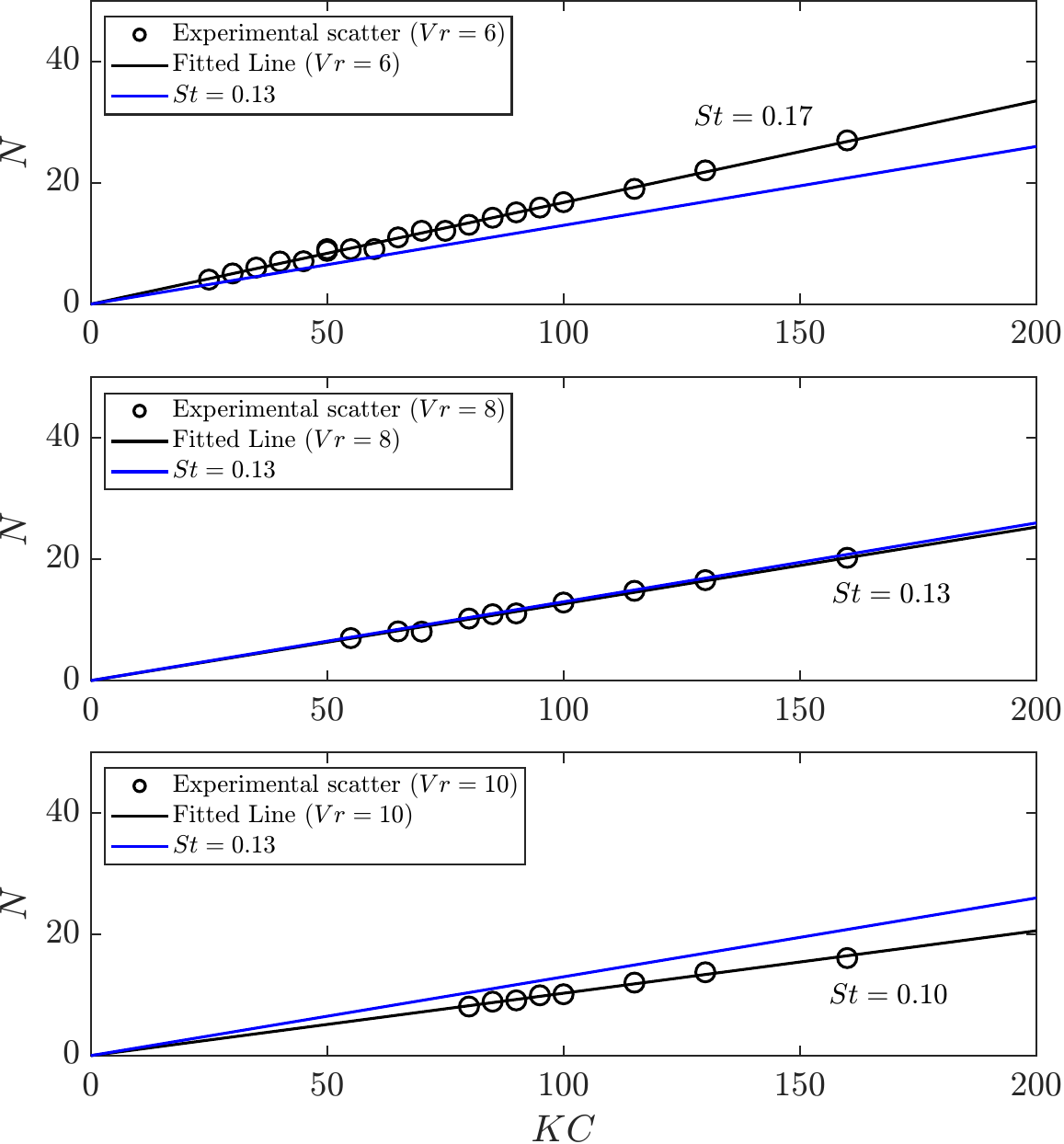}
	\caption{Number of vortex shedding versus KC number for different cases: (a) $Vr=6$, (b) $Vr=8$, and (c) $Vr=10$. Black circle: results of the different cases obtained from the experiment; black line: fitted line based on the experimental data. Blue line: curve with the slope ($St$) of $0.13$.}
	\label{NversusKC}
\end{figure}
}

\begin{figure}[ht!]
	\centering
	\includegraphics[width=.6\textwidth]{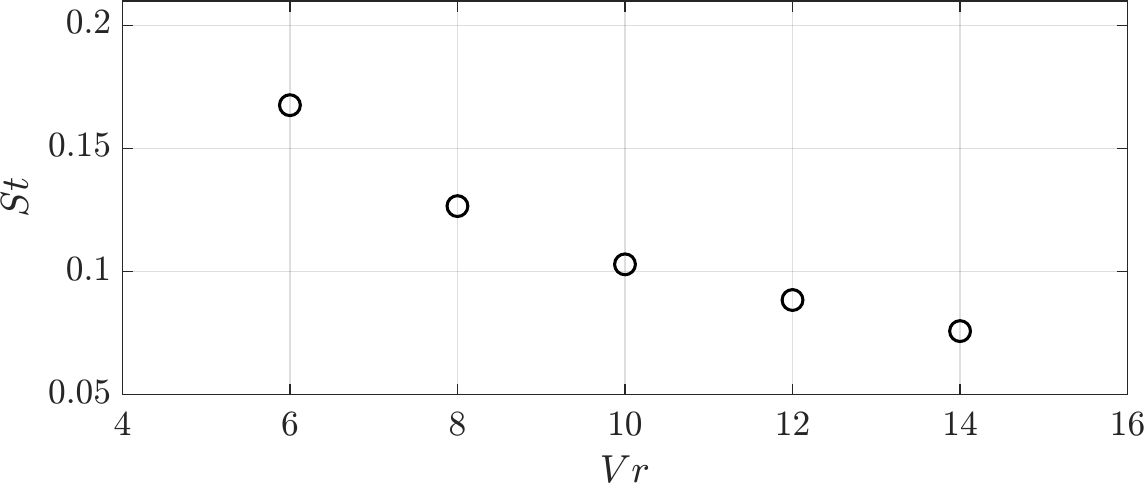}
	\caption{Summarized relationship of the fitted $St$ results with different $Vr$.}
	\label{vrversusst}
\end{figure}

\cref{vrversusst} represents the $St$ result with different $Vr$. The results show that $St$ basically decreases with an increasing $Vr$ in the present study. It should be noted that in the oscillatory sheared flow test represented in the present study, as well as the previously investigated oscillatory uniform flow test, the VIV response is dominated by the first order in most cases. This Strouhal number law reported in \cref{vrversusst} is significantly influenced by the first-order dominant VIV. Future work will cover the experimental study of oscillatory flow-induced high-mode VIV.

\subsection{VIV developing process in oscillatory sheared flow}

\begin{figure}[ht!]
	\centering
	\includegraphics[width=.6\textwidth]{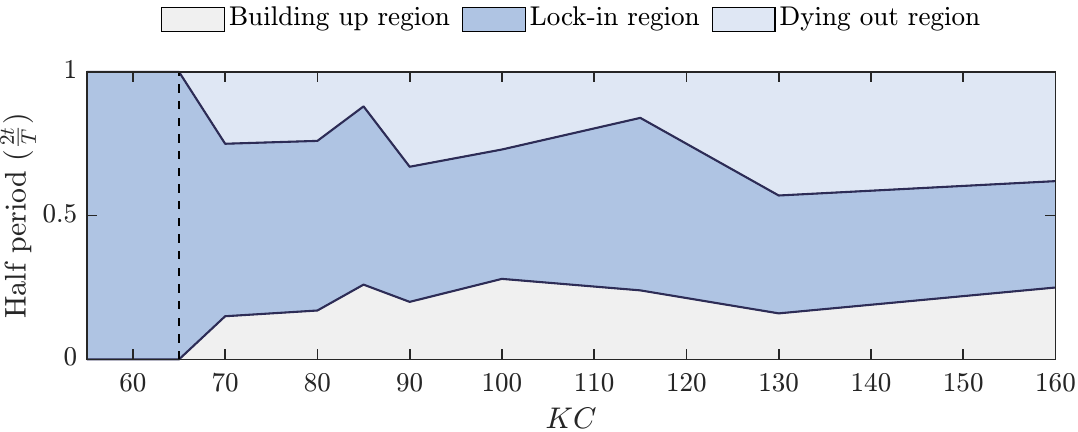}
	\caption{Time interval distribution of the VIV developing process with $Vr = 8 $. Gray area: building up region, deep blue: lock-in region, light blue: dying out region.}
	\label{Vr8internal}
\end{figure}

\begin{figure}[ht!]
	\centering
	\includegraphics[width=.6\textwidth]{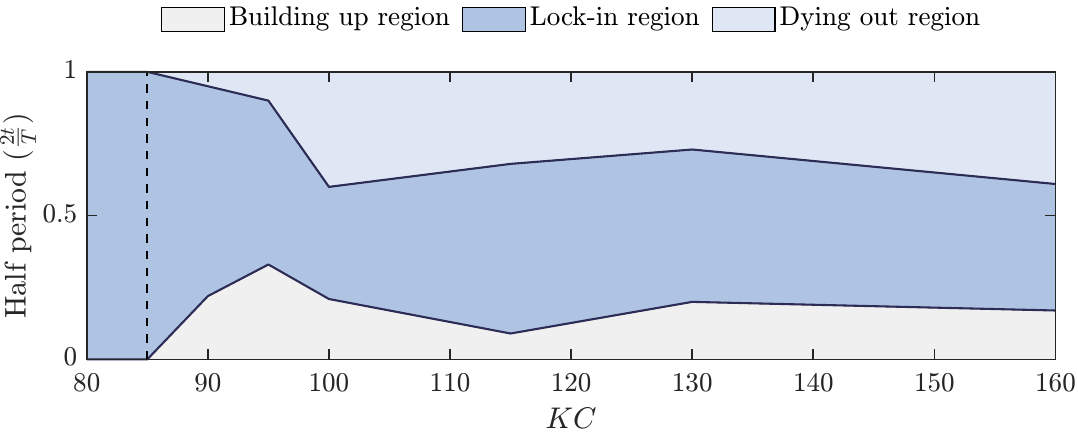}
	\caption{Time interval distribution of the VIV developing process with $Vr = 10 $. Gray area: building up region, deep blue: lock-in region, light blue: dying out region.}
	\label{Vr10internal}
\end{figure}

As \cref{vr8kc160time} states, there exist three regions in the developing process in oscillatory sheared flow. We use the value of 
$(A_y/D)_{max}/\sqrt{2}$ to determine the three regions quantitatively. \cref{Vr8internal} and \cref{Vr10internal} display the time interval distribution of the VIV developing process in the cases with $Vr=8$ and $Vr=10$, respectively. 

There exists a critical $KC$. The response with the $KC$ smaller than the critical $KC$ is steady, which is similar to the behavior of steady flow-induced VIVs. When $KC$ is larger than the critical $KC$, a modulation response occurs with three developing regions. The critical $KC$ in the case with $Vr=8, 10$ are $65$ and $85$. The critical $KC$ increases as the reduced velocity increases. With a further increase in $KC$ larger than $100$, the VIV development process stabilizes with an approximately $50\%$ lock-in region. The critical $KC$ is related the $\gamma$ parameter, further mechanism studies about will critical $KC$ and $\gamma$ be carried out in the future.

\section{Summary}
In the present study, we conducted an experiment with a flexible pipe that is $\SI{28.41}{mm}$ in diameter and $\SI{3.88}{m}$ in length. One end of the test pipe is fixed, and one end is forced to harmonically oscillate to simulate oscillatory sheared flow with various combinations of amplitudes and periods, with KC numbers from $25$ to $160$, and with five kinds of reduced velocities $Vr$ from $6$ to $14$.

There exists obvious modulated response in oscillatory sheared flow induced VIV with building up region, lock-in region and dying out region at high $KC$. As $KC$ increases, the lock-in region reaches a constant time interval distribution of nearly $50\%$. Additionally, hysteresis phenomenon occurs in oscillatory sheared flow induced VIV with a relative higher response amplitude and smaller gradient in the deceleration region.

The response frequency represents multi-frequency with forced vibration frequency harmonics at low $KC$ cases and broadband response at large $KC$ cases. The Strouhal number in oscillatory sheared flow induced VIV is relatively smaller than that in oscillatory uniform flow and decreases as the reduced velocity increases.

The $St$ and critical $KC$ proposed in the present study will be the input for the VIV prediction method for engineering applicant and corresponding research will be conducted soon.

\section*{Appendix A: Mechanism of hysteresis} \label{appendixa}
\begin{figure}[ht!]
	\centering
	\includegraphics[width=.45\textwidth]{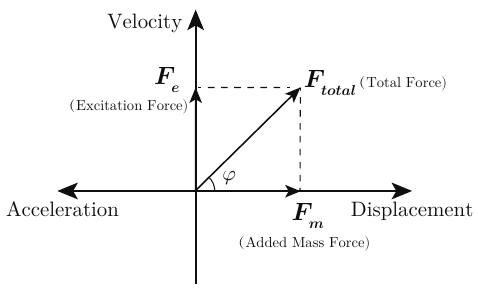}
	\caption{Diagram of the force decomposition method. $\varphi$ is the phase difference between the total lift force and the displacement \citep{fu2022frequency}.}
	\label{forcedecom}
\end{figure}

In this appendix, the mechanism of hysteresis is investigated briefly by the VIV force identification theory and forgetting factor least squares algorithm (FFLS). The VIV force of the pipe can be identified by Euler-Bernoulli beam theory as:
\begin{equation}
        EI\frac{\partial^4y(z,t)}{\partial z^4}-\frac\partial{\partial z}{\left[T(t)\frac{\partial y(z,t)}{\partial z}\right]}+c\frac{\partial y(z,t)}{\partial t}+\bar{m}\frac{\partial^2y(z,t)}{\partial t^2}=F_{total}(z,t),
\end{equation}
where $EI$ is the bending stiffness, $\bar{m}$ is the mass per unit length in air, $T(t)$ is the time-varying tension, $c=2\bar{m}\omega\zeta$ is structural damping, $\omega$ is the dominant circular frequency, $\zeta$ is damping ratio as \cref{tab1} shows. The VIV force can be decomposed as excitation force $F_e$ in phase with velocity and added mass force $F_m$ in phase with acceleration as:
\begin{equation}\label{forcedeeq}
    F_{total}(t)=\underbrace{\frac{\dot{y}(t)}{2|\dot{y}(t)|}\rho DV(t)^2C_e(t)}_{F_e}\underbrace{-\vphantom{\frac{1}{2\sqrt{2}\dot{y}_{rms}}}\frac{\pi D^2}{4}\rho C_m(t)\ddot{y}(t)}_{F_m},
\end{equation}
where $C_e$ is the excitation coefficient, $C_m$ is the added mass coefficient, $V$ is the flow velocity at $z$ location, and the diagram of the force decomposition method is shown in \cref{forcedecom}.

Then the FFLS algorithm is applied for the identification of the time-varying VIV force coefficients, the \cref{forcedeeq} can be rewritten as \citep{fu2022frequency,liu2018time}:
\begin{equation}
\begin{aligned}
    \mathbf{F}_{s}^*&=\mathbf{H}_{s}^*\boldsymbol{\theta}_s,\\
    \mathbf{F}_{s}^*&=\left[\beta^{s-1}F_{total}\left(t_{1}\right), \beta^{s-2}F_{total}^*\left(t_{2}\right), \beta^{s-3}F_{total}\left(t_{3}\right), \ldots,\beta^0 F_{total}\left(t_{s}\right)\right]^{T},\\
    \mathbf{H}_{s}^*&=[\beta^{s-1}\boldsymbol{h}(1), \beta^{s-2}\boldsymbol{h}(2), \beta^{s-3}\boldsymbol{h}(3),\ldots, \beta^{0}\boldsymbol{h}(s)]^{T}\\
    &=\left[\begin{array}{l}
        \dot{y}\left( t_{1}\right), \dot{y}\left(t_{2}\right), \dot{y}\left( t_{3}\right), \ldots, \dot{y}\left(t_{s}\right) \\
        \ddot{y}\left( t_{1}\right), \ddot{y}\left(t_{2}\right), \ddot{y}\left( t_{3}\right), \ldots, \ddot{y}\left(t_{s}\right)
    \end{array}\right]^{T},\\
    \boldsymbol{\theta}(\mathrm{s})&=\left[\begin{array}{c}
        \frac{\rho D V(t)^{2}}{2|\dot{y}(t)|} C_e\left(t_{s}\right) \\
        -\frac{\rho \pi D^{2}}{4} C_m\left(t_{s}\right)
    \end{array}\right],\\
    s&\in \mathbb{N}^+,
\end{aligned}
\label{forcematrixFFLS}
\end{equation}
where $\beta\in(0,1]$ represents the forgetting factor, $\beta=1.00$ in the least squares algorithm. The parameters to be identified need to minimize the sum of the squared errors between $\mathbf{H}_{s}^*\boldsymbol{\theta}_s$ and $\mathbf{F}_{total}^*$ as follows:
\begin{equation}
    \left.\dfrac{\partial J(\boldsymbol{\theta})}{\partial \boldsymbol{\theta}}\right|_{\widehat{\boldsymbol{\theta}}_s}=\left.\dfrac{\partial}{\partial \boldsymbol{\theta}}\left(\mathbf{F}_{s}^{*}-\mathbf{H}_{s}^{*} \boldsymbol{\theta}_s\right)^{T}\left(\mathbf{F}_{s}^{*}-\mathbf{H}_{s}^{*} \boldsymbol{\theta}_s\right)\right|_{\widehat{\boldsymbol{\theta}}_s}=\boldsymbol{0},
\end{equation}
then we can have:
\begin{equation}
    \widehat{\boldsymbol{\theta}}_s=\left(\mathbf{H}_{s}^{*T} \mathbf{H}_{s}^{*}\right)^{-1} \mathbf{H}_{s}^{*T} \mathbf{F}_{s}^{*}.
    \label{FFLSerror}
\end{equation}
	
Substituting \cref{forcematrixFFLS} into \cref{FFLSerror}, we can finally obtain
\begin{equation}
\begin{aligned}
    \widehat{\boldsymbol{\theta}}_s &=\left[\sum_{i=1}^{s} \beta^{2(s-i)} \boldsymbol{h}(i) \boldsymbol{h}^{T}(i)\right]^{-1}\left[\sum_{i=1}^{s} \beta^{2(s-i)} \boldsymbol{h}(i) {F}_{total}(i)\right] \\
    &=\left[\sum_{i=1}^{s} \mu^{(s-i)} \boldsymbol{h}(i) \boldsymbol{h}^{T}(i)\right]^{-1}\left[\sum_{i=1}^{s} \mu^{(s-i)} \boldsymbol{h}(i) {F}_{total}(i)\right] \\
    &=\left(\boldsymbol{H}_{s}^{T} \mathbf{\Lambda}_{s} \boldsymbol{H}_{s}\right)^{-1} \boldsymbol{H}_{s}{ }^{T} \boldsymbol{\Lambda}_{s} \boldsymbol{F}_{s}
\end{aligned}
\end{equation}
where $\mu = \beta^2, \mu\in(0,1]$, $\mu$ is called the forgetting factor, and $\Lambda_{s}$ is the weighted matrix, which is a diagonal matrix $\Lambda_{s}=diag (\mu^{s-1},\mu^{s-2},\ldots,\mu,1)$. The essence of this method is to give different weights to the data depending on the moment of data.

\begin{figure}[ht!]
	\centering
	\includegraphics[width=.6\textwidth]{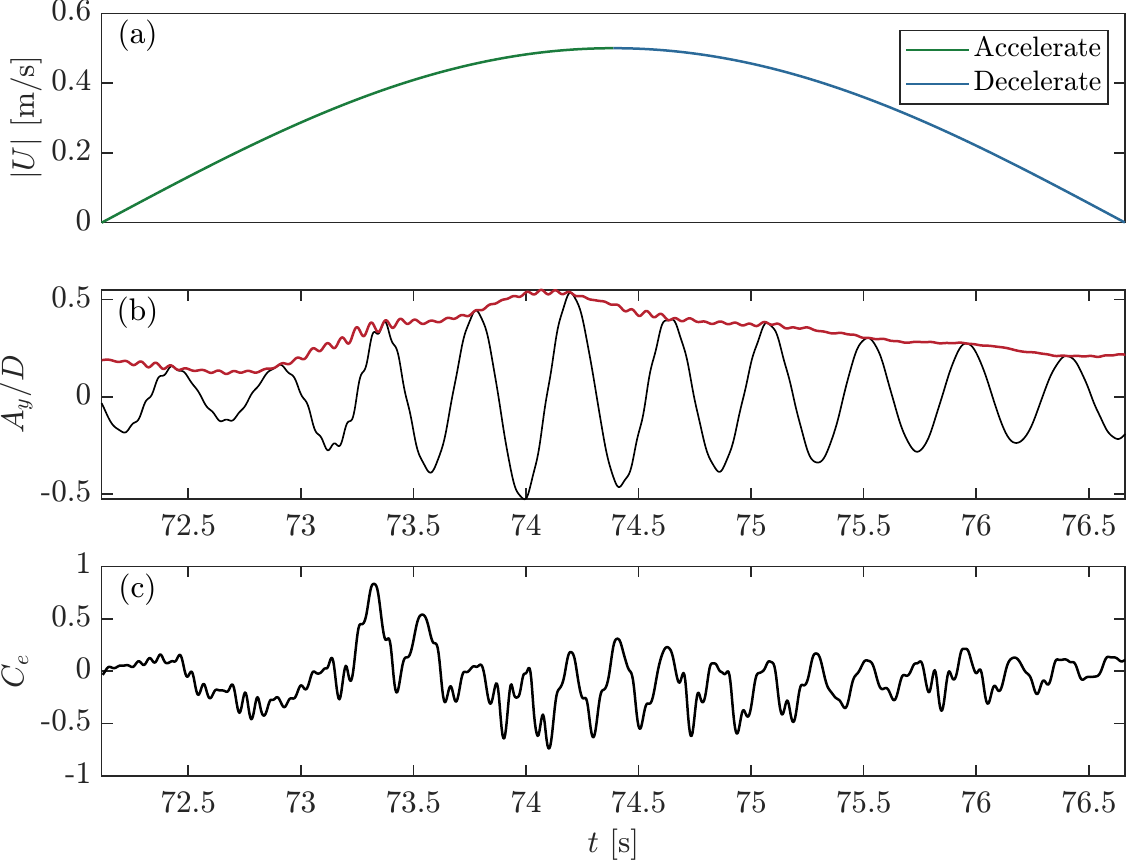}
	\caption{VIV hydrodynamic of in case $Vr=8, KC=160$ at FBG-CF3 point: (a) absolute value of relative flow velocity; (b) VIV displacement; (c) time-varying excitation coefficient $C_e$ at FBG-CF3 point.}
	\label{hysana}
\end{figure}

Excitation coefficient $Ce$ represents the energy transfer between fluid and structure. A positive $C_e$ means that energy transfer from fluid to structure. \cref{hysana} displays the time-varying response and excitation coefficient of half circle. In the acceleration region, there exists a positive peak value of excitation coefficient $C_e$. The $C_e$ exhibit periodic variations but do not show distinct positive peaks in deceleration region. This implies that during the acceleration region, compared to the deceleration phase, the flexible pipe extracts more energy from the fluid to support the vibration which resulted in a higher gradient. More detailed fluid-structure interaction mechanism analysis will be conducted based on further CFD research.

\section*{Acknowledgments}
The authors gratefully acknowledge the financial support from the National Natural Science Foundation of China under Grant Number of  52001208 and 52088102, the Fundamental Research Funds for the Central Universities, the Joint Funds of the National Natural Science Foundation of China under Grant Number of U19B2013, Shanghai Science and Technology Program under Grant Number of 21ZR1434500, 22ZR1434100, National Science Fund for Distinguished Young Scholars under Grant Number of 51825903, State Key Laboratory of Ocean Engineering (Shanghai Jiao Tong University) under Grant Number of GKZD010081, Shenlan Project under Grant Number of SL2021MS018, Young Elite Scientists Sponsorship Program under Grant Number of 2020QNRC001. The authors would also like to express gratitude to DNV for preparing experiment.

%

\end{document}